\documentclass[aoas,preprint]{imsart}
\RequirePackage[numbers]{natbib}
\usepackage{geometry} 
\usepackage{amssymb}
\RequirePackage[OT1]{fontenc}
\RequirePackage{amsthm,amsmath}
\usepackage{tabulary}
\usepackage{multirow}
\usepackage{color}
\usepackage{xcolor,colortbl}
\usepackage{bbm}
\usepackage{graphicx}
\usepackage{epstopdf}
\usepackage{float}

\newcommand\independent{\protect\mathpalette{\protect\independenT}{\perp}}
\def\independenT#1#2{\mathrel{\rlap{$#1#2$}\mkern2mu{#1#2}}}

\arxiv{arXiv:0000.0000}

\startlocaldefs
\numberwithin{equation}{section}
\theoremstyle{plain}

\endlocaldefs

\begin{document}

\begin{frontmatter}
\title{A Note on Estimating Optimal Dynamic Treatment Strategies Under Resource Constraints Using Dynamic Marginal Structural Models}
\runtitle{Optimal Resource Constrained Dynamic Strategies}

\begin{aug}
\author{\fnms{Ellen C} \snm{Caniglia}\thanksref{m1,m2}\ead[label=e1]{ellen.caniglia@nyulangone.edu}},
\author{\fnms{Eleanor J} \snm{Murray}\thanksref{m2}\ead[label=e2]{emurray@mail.harvard.edu}},
\author{\fnms{Miguel A} \snm{Hern\'an}\thanksref{m2}\ead[label=e3]{mhernan@hsph.harvard.edu}}
\and
\author{\fnms{Zach} \snm{Shahn}\thanksref{m3}\ead[label=e4]{zach.shahn@ibm.com}}
\runauthor{E. Caniglia et al.}

\affiliation{New York University School of Medicine\thanksmark{m1}, Harvard T. H. Chan School of Public Health\thanksmark{m2} and IBM Research \thanksmark{m3}}

\address{New York University Medical School \\
Department of Population Health \\
227 E. 30th St, New York, NY, USA, 10016  \\
\printead{e1} 
\phantom{E-mail: ellen.caniglia@nyulangone.edu\ }}

\address{Harvard T.H. Chan School of Public Health \\
Department of Epidemiology \\
677 Huntington Avenue, Boston, MA, USA, 02115\\}

\address{IBM Research\\
1101 Kitchawan Road, Yorktown Heights, NY, USA, 10598}
\end{aug}

\begin{abstract}
Existing strategies for determining the optimal treatment or monitoring strategy typically assume unlimited access to resources. However, when a health system has resource constraints, such as limited funds, access to medication, or monitoring capabilities, medical decisions must balance impacts on both individual and population health outcomes. That is, decisions should account for competition between individuals in resource usage. One simple solution is to estimate the (counterfactual) resource usage under the possible interventions and choose the optimal strategy for which resource usage is within acceptable limits. We propose a method to identify the optimal dynamic intervention strategy that leads to the best expected health outcome accounting for a health system's resource constraints. We then apply this method to determine the optimal dynamic monitoring strategy for people living with HIV when resource limits on monitoring exist using observational data from the HIV-CAUSAL Collaboration. 

\end{abstract}



\end{frontmatter}

\section{Introduction}
Physicians repeatedly assess patients with chronic health conditions and make treatment decisions based on patient history at each assessment. Here `treatment' refers to any intervention, including monitoring through lab tests to inform future decisions. A `dynamic treatment strategy' is a function mapping a patient's treatment and covariate history up to the current visit to a treatment decision at that visit.  

In resource limited settings, where health system constraints prevent immediate initiation of treatment in all individuals, the optimal dynamic strategy is the strategy which, if implemented by all doctors, would lead to the best population health outcomes while `respecting' the system's resource constraints (in a sense we will make rigorous below). While randomized clinical trials of a wide range of dynamic strategies may be ideal for estimating this optimal strategy, they are usually financially or logistically infeasible. However, observational data can be used to estimate the optimal dynamic strategy under resource constraints. Note that, as would be the case with randomized trials, the optimal strategy here refers to the optimum from among a class of options assessed -- the true overall optimum may not be among these.

Luedtke and van der Laan \cite{r1} considered this problem for the case of a point exposure -- that is, when there is only one time point at which a treatment decision is made. They considered resource constraints that place an upper limit $\kappa$ on the  expected proportion of treated patients, and defined the set of strategies which respected resource constraint $\kappa$ as all strategies for which the expected proportion of treated patients in the population under that strategy is less than $\kappa$. The optimal point exposure strategy was then identified as the optimal strategy among those respecting the system constraint. However, optimal point exposure strategies are often of limited utility in clinical decision-making, especially in the context of chronic disease or long-term therapy. Such strategies cannot recommend, for example, ``\emph{Come back next month, and if the problem has progressed then begin treatment}''. 

Here, we consider the optimal resource constrained dynamic strategy (RCDS) and present a method for estimating the optimal RCDS from a parameterized subclass of all strategies. For example, suppose we restrict our attention to the class of monitoring strategies $\{g_x: g_x(\bar{L}_{t-1},\bar{N}_{t-1}) = \mathbf{1}\{L_{t-1}>x\}\}$ (where $\mathbf{1}\{ \}$ denotes the indicator function) that monitor ($N_{t}$) at time $t$ if and only if the covariate $L_{t-1}$ is greater than $x$. Such a class of strategies might be approximately appropriate, for example, for the decision of how often to monitor for anti-retroviral therapy (ART) failure or resistance in people living with HIV, where $L_t$ represents CD4 count at time $t$.

In the absence of resource constraints, Orellana et al. \cite{r2} and Robins et al. \cite{r3} describe how to estimate the optimal strategy from a parameterized class of strategies using a dynamic Marginal Structural Model (dyn-MSM). A dyn-MSM models the expected counterfactual outcomes under strategies parameterized by $x$ as a function $b(x;\beta)$ of $x$. With an estimate $\hat{\beta}$ of the dyn-MSM parameter $\beta$, estimating the optimal strategy simply reduces to finding $x^{opt}$ maximizing $b(x;\hat{\beta})$ (assuming larger values of $b(x;\beta)$ are preferable). To accommodate resource constraints, we propose a fairly straightforward extension of this procedure that entails fitting two dyn-MSMs--one estimating the expected counterfactual clinical outcome $b(x;\beta)$ and one estimating expected counterfactual treatment utilization $h(x;\theta)$. The optimal RCDS is then $x^{opt}_{rc}$ maximizing  $b(x;\beta)$ subject to $h(x;\theta)<\kappa$.

We apply this approach to estimate the optimal RCDS for CD4 cell count and HIV-RNA monitoring in people living with HIV who have achieved viral suppression. CD4 cell count and HIV-RNA tests are used to monitor an individual's response to ART. More frequent monitoring has been shown to be associated with a lower risk of virologic failure (HIV RNA levels $>200$ copies/ml) at two years after viral suppression \cite{r4}. Guidelines recommend dynamic monitoring strategies in which virologically suppressed individuals on ART may be monitored less frequently once their CD4 cell count crosses above a certain threshold. However, the optimal point at which to decrease monitoring frequency is unclear even for high-income settings. In a health system with limited funds for monitoring, the CD4 cell count at which monitoring may be decreased must be chosen from a subset of strategies where the average number of tests does not exceed the available resources.

\section{Notation}
\noindent Let:
\begin{itemize}
\item$t \in \{0,\ldots,K\}$ denote time, assumed discrete, with $K$ the end of the study;
\item $N_t$ be a variable indicating the whether an individual is monitored at time $t$;
\item$Y$ denote health outcome we aim to optimize;
\item$L_t$ denote covariates at time $t$, including past treatments, that may influence monitoring decisions; 
\item $D$ denote the total number of monitoring tests received by an individual;

\item $\bar{X}_t$ denote $X_0,\ldots,X_t$ and $\underline{X}_t$ denote $X_t,\ldots,X_K$ for arbitrary time varying variable $X$.
\end{itemize}

We assume that we observe $iid$ realizations of the random vector \\
$\mathbbm{O}=(\textbf{L}_0,\ldots,\textbf{L}_K,N_0,\ldots,N_K,Y)$. We use capital letters to denote random variables and corresponding lower case letters to denote specific values that random variables might take.

A treatment $strategy$ is a rule or function that determines the value to which $N_t$ will be set for a given observed history, i.e. a function $g:(\bar{L}_{t-1},\bar{N}_{t-1})\mapsto N_t$. A strategy is said to be $static$ if its recommendation for the present does not depend on past covariate and treatment values, and can therefore be specified from baseline. An example of a static strategy would be `monitor every 3 months'. A strategy is said to be $dynamic$ if it does depend on past covariate and treatment values. An example of a dynamic strategy would be `monitor if time since last monitoring $\geq$ 6 months or if last observed viral load $>$ 100 copies/ml and time since last monitoring $\geq$ 2 months'. Most realistic, clinically-relevant strategies are dynamic.

We denote arbitrary strategies by $g$ and we adopt the counterfactual framework of Robins \cite{r5} in which corresponding to each possible strategy $g$ are counterfactual random variables $Y(g)$, $L_t(g)$, and $D(g)$ that would have been observed had strategy $g$ been followed, possibly contrary to fact. Implicit in the notation for counterfactuals (e.g. $Y(g)$) is the assumption that the treatment strategy followed by one patient does not influence any other patient. This implicit assumption is called the `No Interference Assumption' by Rubin \cite{r6}. Note that in defining our resource constraint, we are optimizing based on the (counterfactual) average number of treatment or monitoring events per individual over a defined time period, but assuming no competition between individuals to access care under the optimal RCDS. We also make the additional assumptions \cite{r7}:

\begin{align}
\begin{split}\label{positivity}
&\textbf{Sequential Positivity} \text{: If } f(\bar{l}_{t},\bar{n}_{t-1})>0,\text{ } f(N_t=n|\bar{L}_{t}=\bar{l}_{t},\bar{N}_{t-1}=\bar{n}_{t-1}) > 0 \text{ }\forall n,t
\end{split}\\
\begin{split}\label{consistency}
&\textbf{Consistency} \text{: For any strategy }g, \text{ if for a given subject } N_t=g(\bar{L}_t,\bar{N}_{t-1}) \text{ for each }t,\\
&\text{ then } Y = Y(g) \text{ and } \bar{L}_K=\bar{L}_K(g)\text{ for that subject}
\end{split}\\
\begin{split}\label{seq_ex}
&\textbf{Sequential Exchangeability}\text{: } Y(g) \independent N_t | \bar{L}_{t-1}=\bar{l}_t,\bar{N}_{t-1}=\bar{n}_{t-1} \text{ } \forall t, g,\bar{n},\bar{l}
\end{split}
\end{align}

We define resource constraints as caps on the expected number of doses or monitoring events per patient over a defined time period. We say that strategy $g$ respects resource constraint $\kappa$ if
\begin{equation}
E[D(g)] < \kappa
\end{equation}
We consider parameterized classes of strategies $\{g_x\}$ and seek to estimate 
\begin{equation}
x_{\kappa}^{opt} \equiv argmax_x E[Y(g_x)] \text{ subject to } E[D(g_x)]<\kappa
\end{equation}

\section{Review of dyn-MSMs}
A dyn-MSM is a model for expected counterfactual outcomes under a class of strategies parameterized by $x$ as a function of $x$, i.e.
\begin{equation}
E[Y(g_x)] = b(x;\beta)
\end{equation}
To estimate $\beta$, first note that each subject might follow multiple strategies from the class $\{g_x\}$. Let $\Lambda_i$ denote the number of strategies that subject $i$ follows. Generate an artificial dataset with $\Lambda_i$ contributions from each subject: $(Y_i,x_{i1}),\ldots,(Y_i,x_{i\Lambda_i})$. Using the artificial dataset, we can fit by weighted least squares the regression model 
\begin{equation*}
E[Y|x] = b(x;\gamma)
\end{equation*}
with weights
\begin{equation*}
W(x)\equiv \prod_{k=0}^K\frac{f^*(x)}{f(N_k|\bar{L}_{k-1},\bar{N}_{k-1})}
\end{equation*}
to obtain $\hat{\gamma}$. When treatment or monitoring probabilities are unknown, as they are in our application, a consistent estimator $\hat{f}(N_k|\bar{L}_{k-1},\bar{N}_{k-1})$ of $f(N_k|\bar{L}_{k-1},\bar{N}_{k-1})$ can be plugged into $W(x)$. Under sequential exchangeability (\ref{seq_ex}) and consistency (\ref{consistency}), \cite{r2} shows that the weighted regression parameter estimate $\hat{\gamma}$ approaches the causal estimand $\beta$ in the limit. 

\section{Estimating Optimal Treatment Strategies With Resource Constraints} 
To estimate the optimal RCDS, we simply estimate the parameters of two dyn-MSMs:
\begin{align}
E[Y(g_x)] = b(x;\beta)\\
E[D(g_x)] = h(x;\theta)
\end{align}
and then estimate the $x$ indexing the optimal strategy as
\begin{align*}
\begin{split}
\hat{x}^{opt}_{\kappa} \equiv argmax_x\text{ } b(x;\hat{\beta})\\
\text{subject to } h(x;\hat{\theta}) \leq \kappa
\end{split}
\end{align*}
Standard errors for $\hat{\beta}$, $\hat{\theta}$, and certain derived quantities can be computed by bootstrap or analytically using formulas in \cite{r2}. 
\section{Application to Monitoring of HIV Patients}
We apply the method described above to estimate the optimal RCDS for monitoring CD4 cell count and HIV-RNA in people living with HIV using data from the HIV-CAUSAL collaboration. The HIV-CAUSAL collaboration combines data from prospective cohorts of people living with HIV enrolled in universal health care systems in Brazil, Canada, France, Greece, Netherlands, Spain, Switzerland, UK, and USA. 

We have previously reported on the optimal dynamic monitoring strategy in this cohort and showed that decreasing monitoring when CD4 cell count $>$200 cells/$\mu$l compared to $>$500 cells/$\mu$l does not worsen short-term clinical and immunologic outcomes in virologically suppressed individuals living with HIV but may increase the risk of virological failure  \cite{r4}.  We now extend these results to identify the optimal RCDS under constraints on the average number of monitoring events over a two-year period. However, because the majority of the HIV-CAUSAL data comes from high-income countries, we apply an artificial resource constraint selected to demonstrate the methodologic approach. 

First we briefly describe the eligibility criteria and monitoring strategies under consideration. We then describe the estimation of the optimal RCDS.

\emph{Eligibility criteria}: Previously antiretroviral therapy naive HIV-positive individuals who initiate antiretroviral therapy in 2000 or later and achieve confirmed virologic suppression (2 consecutive HIV-RNA $\le$200 copies/ml) within 12 months of initiating therapy are eligible for inclusion in the study. Individuals must meet the following additional eligibility criteria at baseline (date of confirmed virologic suppression): 18 years of age or older, CD4 cell count measurement within the previous 3 months, no history of an AIDS-defining illness, and no pregnancy (when information was available). 

\emph{Monitoring strategies}: We consider 31 dynamic monitoring strategies, based loosely on current clinical guidelines. Under each strategy, CD4 cell count and HIV-RNA are monitored every 3-6 months when CD4 is below the strategy's threshold and every 9-12 months when CD4 is above the threshold. Each strategy corresponds to a CD4 threshold ranging from 200-500 cells/$\mu$l in increments of 10 cells/$\mu$l. All of the monitoring strategies further require individuals to be monitored once every 3-6 months when HIV-RNA $>$200 copies/ml or after diagnosis of an AIDS-defining illness, and that CD4 cell count and HIV-RNA be monitored concurrently. 

\emph{Follow-up period and outcome}: Individuals are followed from baseline until death, pregnancy (if known), loss to follow-up, or the administrative end of follow-up. The outcome of interest is virologic failure (HIV-RNA $>$200 copies/ml) at 24 months of follow-up. 

\emph{Statistical methods}: We compare the 31 monitoring strategies using the replication and censoring approach. Briefly, we create an expanded dataset by making 31 exact replicates of each individual (1 per strategy). If and when an individual's data are no longer consistent with a given strategy, we artificially censor the corresponding replicate at that time. We compute inverse probability weights to adjust for the potential selection bias induced by the artificial censoring.

We then fit an inverse-probability weighted Poisson regression model to estimate the risk ratio of virologic failure at 24 months of follow-up among those with measurements at 24 $\pm$ 2 months. The model includes a flexible functional form of the strategy variable (restricted cubic splines) and the baseline covariates: sex, CD4 cell count ($<$200, 200-349, 350-499, $\geq$500 cells/μL), years since HIV diagnosis ($<$1, 1 to 4, $\geq$5 years, unknown), race (white, black, other or unknown), geographic origin (N. America/W. Europe, Sub-Saharan Africa, other, unknown), acquisition group (heterosexual, homosexual or bisexual, injection drug use, other or unknown), calendar year (restricted cubic splines with 3 knots at 2001, 2007, and 2011), age (restricted cubic splines with 3 knots at 25, 39, and 60), cohort, and months from cART initiation to virologic suppression (2-4, 5-8, $\geq$9). Under the assumptions described above, the parameters of the regression model consistently estimate the parameters of a dynamic marginal structural model. The model's estimated parameters are used to estimate the standardized risk of virologic failure at 24-months for each monitoring strategy.

Next, we fit an inverse-probability weighted log-linear regression model to estimate the mean number of measurements at 24 months of follow-up. As above, the model includes a flexible functional form of the strategy variable and the baseline covariates. The predicted values are used to estimate the standardized mean number of measurements at 24-months for each monitoring strategy.  

After estimating the counterfactual risk of virologic failure at 24 months and counterfactual mean number of measurements at 24 months, we rank the strategies by the counterfactual mean number of measurements at 24 months. We then restrict our consideration to the strategies that satisfy the resource constraint. For our example, we consider a hypothetical constraint allowing an average of one CD4 cell count and one HIV-RNA test per person every 6 months, for an average of 4 measurements per person over 24 months. Under this constraint, only the strategies that lead to an average of 4 measurements per person over the 24-months of follow-up will be considered. 

Finally, we find the optimal strategy for minimizing the risk of virologic failure among the strategies that satisfy the constraint. 

\emph{Results}: Figure 1 shows the estimates obtained from the dyn-MSMs for virologic failure at 24 months and for mean number of monitoring events over 24 months across the range of CD4 cell count thresholds considered. In this example, the estimated risk of virologic failure is monotonically increasing  and the estimated mean number of measurements is monotonically decreasing as the CD4 cell count threshold increases, so the optimal RCDS can be identified graphically as the lowest CD4 cell count threshold for which the mean number of monitoring events over 24 months is below the resource constraint, $\kappa$. Table 1 gives the same information with 95\% confidence intervals obtained via 500 bootstrap samples.

In our example, we consider the case of $\kappa$ = 4 and identify the optimal threshold for switching monitoring frequency as 320 cells/$\mu$l. The optimal RCDS is then `monitor CD4 cell count and HIV-RNA every 3-6 months when CD4 is below  320 cells/$\mu$l and every 9-12 months when CD4 is above  320 cells/$\mu$l'.
\newpage
\begin{figure}[h!]
  \caption{Risk of virologic failure and mean number of measurements per person at 24 months of follow-up by CD4 threshold strategy.}
	\includegraphics[width=\textwidth]{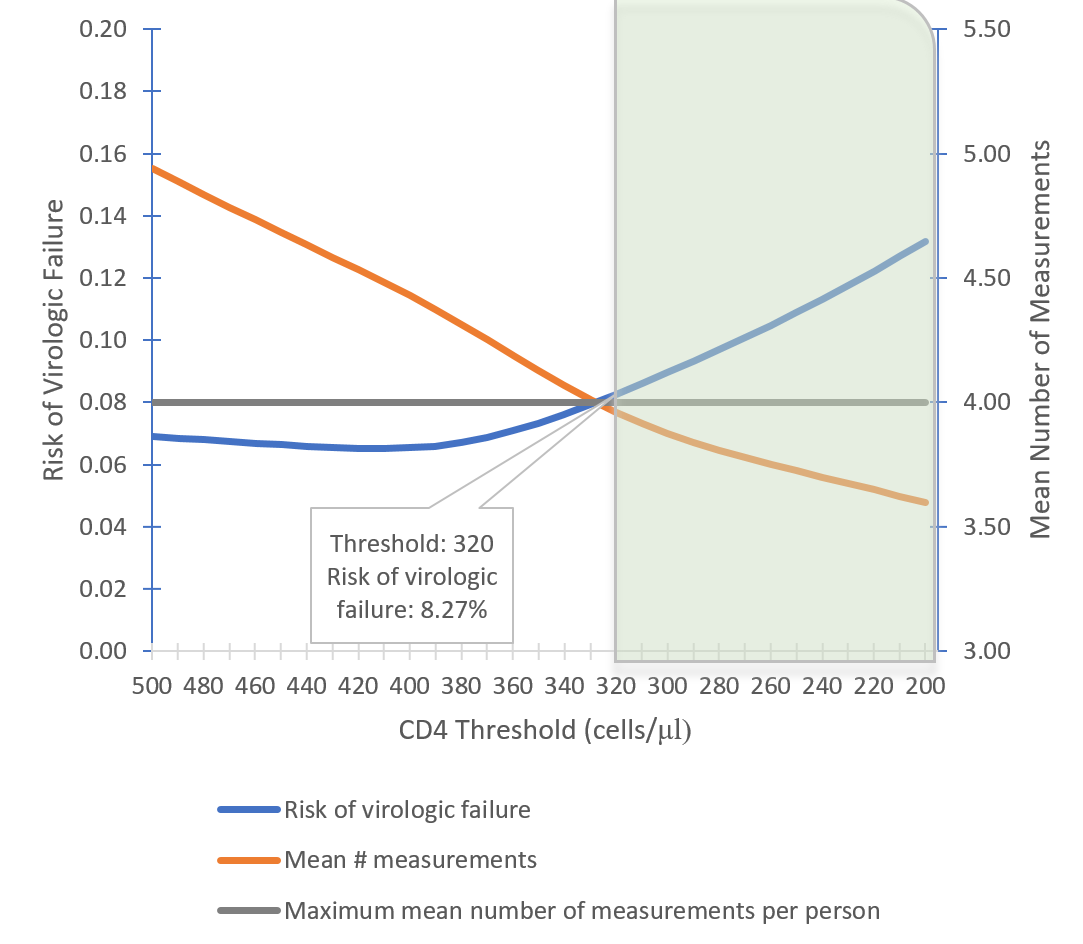}
The grey line represents one potential resource constraint -- a cap on per person number of measurements over 24 months of follow-up. Strategies in the green area meet this restriction, and the CD4 threshold 320 strategy is the optimal RCDS.
\end{figure}

\newpage

\begin{table}[H]
\label{table:table1}
\caption{Risk of Virologic Failure and cumulative number of measurements at 24 months for CD4 threshold} \label{tab:title}
\begin{tabulary}{\linewidth}{lllll}
	\hline
    CD4 Threshold* \\ (cells/$\mu$l) & 
    \multicolumn{2}{l}{Virologic Failure } &  
    \multicolumn{2}{l}{Cumulative \# Measurements} \\
	\hline
     & Risk (\%) & 95\% CI & Expected Value &  95\% CI \\
     \hline
    500 & 6.91  & (3.99, 9.82) & 4.94  & (4.82, 5.06) \\
    490 & 6.85 	& (4.09, 9.61) &	4.89 & (4.78, 5.00)  \\
480 & 6.80  	& (4.17, 9.43) &	4.84 & (4.73, 4.94) \\
470 & 6.74  	& (4.21, 9.27) &	4.78 & (4.69, 4.88) \\
460 & 6.69 	& (4.23, 9.15) &	4.73 & (4.64, 4.83) \\
450 & 6.64 	& (4.22, 9.05) &	4.68 &(4.59, 4.78) \\
440 & 6.59 	& (4.18, 8.99) &	4.63 & (4.54, 4.73) \\
430 & 6.54 	& (4.13, 8.95) &	4.58 & (4.49, 4.68) \\
420 & 6.52 	& (4.08, 8.95) &	4.53 & (4.43, 4.63) \\
410 & 6.51 	& (4.03, 8.99) &	4.48 & (4.38, 4.58) \\
400 & 6.54 	& (4.02, 9.05) &	4.43 & (4.32, 4.53) \\
390 & 6.60 	& (4.05, 9.15) &	4.37 & (4.27, 4.48) \\
380 & 6.71 	& (4.14, 9.28) &	4.31 & (4.21, 4.42) \\
370 & 6.87 	& (4.26, 9.48) &	4.25 & (4.15, 4.36) \\
360 & 7.08 	& (4.41, 9.76) &	4.19 & (4.08, 4.30) \\
350 & 7.33 	& (4.55, 10.12) &	4.13 & (4.02, 4.24) \\
340 & 7.62	& (4.67, 10.57) &	4.07 & (3.95, 4.18) \\
330 & 7.93 & (4.79, 11.08) &    4.01 & (3.89, 4.13) \\
\hline 
\rowcolor{gray!50}
\color{red}320 & \color{red}8.27 & \color{red}(4.91, 11.62) & \color{red}3.96 &\color{red} (3.84, 4.08) \\
\rowcolor{gray!50}
310 & 8.61 	&  (5.09, 12.14) &	3.92& (3.79, 4.04) \\
\rowcolor{gray!50}
300 & 8.97 	& (5.31, 12.61) &	3.88& (3.75, 4.00) \\
\rowcolor{gray!50}
290 & 9.33	& (5.59, 13.06) &	3.84& (3.71, 3.97) \\
\rowcolor{gray!50}
280 & 9.70	& (5.91, 13.49) &	3.81& (3.68, 3.93) \\
\rowcolor{gray!50}
270 & 10.08	& (6.22, 13.94) &	3.78& (3.65, 3.91) \\
\rowcolor{gray!50}
260 & 10.48	& (6.52, 14.43)&	3.75& (3.62, 3.88) \\
\rowcolor{gray!50}
250 & 10.88	& (6.78, 14.99) &	3.73& (3.59, 3.86) \\
\rowcolor{gray!50}
240 & 11.31	& (6.98, 15.64) &	3.70& (3.55, 3.85) \\
\rowcolor{gray!50}
230 & 11.75	& (7.12, 16.38) &	3.67& (3.52, 3.83) \\
\rowcolor{gray!50}
220 & 12.21	& (7.21, 17.72) &	3.65& (3.48, 3.82) \\
\rowcolor{gray!50}
210 & 12.68 & (7.25, 18.12) &	3.62& (3.44, 3.80) \\
\rowcolor{gray!50}
200 & 13.18 & (7.23, 19.13) &	3.60& (3.41, 3.79) \\
\hline \\
\end{tabulary}\\
\small{*The CD4 Threshold corresponds to the CD4 cell count at which monitoring frequency changes from once every 2-7 months (if CD4 cell count is below the threshold) to once every 8-13 months (if CD4 cell count is above the threshold). Each strategy also includes monitoring once every 2-7 months when HIV-RNA\textgreater200 copies/ml or after diagnosis of an AIDS-defining illness.}\\
\small{The monitoring strategies falling in the grey area meet the restriction that CD4 cell count and HIV-RNA may only be monitored every six months. Among the monitoring strategies that meet the restriction, the 320 threshold strategy is the optimal strategy.}
\end{table}

\newpage

\section{Conclusions}

Dynamic treatment strategies are a better representation of real-world clinical decision-making processes than static or point intervention strategies. However, resource utilization of dynamic strategies is difficult to assess, since the number of individuals requiring intervention over time under a given strategy cannot be straightforwardly determined at baseline. When a health system faces resource constraints that prohibit implementing the true optimal dynamic treatment strategy, the optimal RCDS is instead required. 

Here we propose a method to identify the optimal RCDS within a parameterized class of strategies of interest by estimating the counterfactual resource usage. We apply this method to estimate the optimal RCDS for monitoring frequency among individuals living with HIV who achieve virologic suppression.

Our choice of $\kappa$ = 4 was somewhat arbitrary. If we had instead chosen $\kappa$ = 3, we would have found that none of the strategies under consideration would satisfy this resource constraint. Interestingly, if we had chosen $\kappa$ = 4.7, we would have identified the optimal threshold for switching monitoring frequency as 410 cells/$\mu$l, even though all the strategies in the range from 200-450 cells/$\mu$l would have satisfied the resource constraint (however, the confidence intervals around our estimates are quite wide).   

In reality, determining the number of CD4 cell count and HIV-RNA measurements a setting is willing to allocate depends on a complex assessment of the costs and health benefits of monitoring. In our illustrative application, we imposed a constraint on the number of tests, which we imagined was derived from a hypothetical corresponding cost constraint. In other applications, it might be useful to directly bound cost instead. For example, since HIV-RNA tests cost more than CD4 tests, an optimal strategy satisfying a total cost constraint may be a more flexible joint strategy that allows CD4 cell count and HIV-RNA to be monitored with different frequencies. Future studies should also assess other health outcomes such as quality-adjusted life years associated with various monitoring strategies. Finally, even in the absence of a single hard resource constraint, examining outcomes of optimal strategies over a range of hypothetical cost constraints could allow for computation of incremental cost-effectiveness ratios, which could be useful for key stakeholders and decision-makers.

\section{Acknowledgements}
We thank Andrew Phillips, Linda Wittkop, Giota Touloumi, and Hansjakob Furrer for useful comments on an earlier draft of this paper and James Robins for helpful discussions. This work was partially supported by NIH grants R37 AI102634 and T32 AI007433.

\end{document}